\documentclass[conference]{IEEEtran}
\IEEEoverridecommandlockouts
\usepackage{cite}
\usepackage{amsmath,amssymb,amsfonts}
\usepackage{algorithmic}
\usepackage{graphicx}
\usepackage{textcomp}
\usepackage{xcolor}
\usepackage{cancel,ulem}
\def\BibTeX{{\rm B\kern-.05em{\sc i\kern-.025em b}\kern-.08em
    T\kern-.1667em\lower.7ex\hbox{E}\kern-.125emX}}

\usepackage{subfigure}
\usepackage{color}

\usepackage{xcolor}
\usepackage{color}
\ifodd 0
  \newcommand{\rev}[1]{{\color{blue}#1}} 
  \newcommand{\com}[1]{\textbf{\color{red}(COMMENT: #1)}} 
  \newcommand{\Ora}[1]{{\color{orange}#1}}
  \newcommand{\dlt}[1]{\sout{\color{blue}#1}}
  \newcommand{\rpl}[2]{{\sout{\color{red}#1}}{\color{blue}#2}}
  \newcommand{\Oras}[1]{{\color{blue}#1}}
  \newcommand{\changesforfinalver}[1]{{\color{blue}#1}}
\else
  \newcommand{\rev}[1]{#1}
  
  \newcommand{\com}[1]{}
  \newcommand{\Ora}[1]{#1}
  \newcommand{\dlt}[1]{}
  \newcommand{\rpl}[2]{{}{#2}}
  \newcommand{\Oras}[1]{#1}
  \newcommand{\changesforfinalver}[1]{#1}
\fi

\begin{document}

\title{
Semantic Communications with Explicit Semantic Base for Image Transmission
\thanks{This work is supported, in part, by the National Natural Science Foundation of China under Grant 62293485, 
and in part by the Fundamental Research Funds for the Central Universities under Grant 2022RC18. 
 \rev{Corresponding author:} Fengyu Wang, Email: fengyu.wang@bupt.edu.cn.
}
}


\author{
\IEEEauthorblockN{
Yuan Zheng\IEEEauthorrefmark{1},
Fengyu Wang\IEEEauthorrefmark{2},
Wenjun Xu\IEEEauthorrefmark{1},
Miao Pan\IEEEauthorrefmark{3},
and Ping Zhang\IEEEauthorrefmark{1}}
\IEEEauthorblockA{\IEEEauthorrefmark{1}State Key Laboratory of Network and Switching Technology,\\
Beijing University of Posts and Telecommunications, Beijing, 100876, China}
\IEEEauthorblockA{\IEEEauthorrefmark{2}School of Artificial Intelligence, Beijing University of Posts and Telecommunications, Beijing, 100876, China}
\IEEEauthorblockA{\IEEEauthorrefmark{3}Department of Electrical and Computer Engineering, University of Houston, Houston, TX 77004, USA}
}

\maketitle

\begin{abstract}
Semantic communications, aiming at ensuring the successful delivery of the meaning of information, are expected to be one of the potential techniques for the next generation communications.
However, the knowledge forming and synchronizing mechanism that enables semantic communication systems to extract and interpret the semantics of information according to the communication intents is still immature. 
In this paper, we propose a \rev{semantic image transmission framework with explicit semantic base (Seb),}
where Sebs are generated and \rev{employed}
as the knowledge \rev{shared between the transmitter and the receiver} with flexible granularity. 
\rev{To represent images with Sebs, a novel Seb-based reference image generator is proposed to generate Sebs and then decompose the transmitted images.}
To further encode/decode the residual information for precise image reconstruction, a Seb-based image encoder/decoder is proposed.
\rev{The key components of the proposed framework are optimized jointly by end-to-end (E2E) training, where the loss function is \changesforfinalver{dedicatedly} designed to tackle the problem of non-differentiable operation in Seb-based reference image generator by introducing a gradient approximation mechanism.}
\rev{Extensive experiments show that the proposed framework outperforms state-of-art works by $0.5 - 1.5$ dB in peak signal-to-noise ratio (PSNR) w.r.t. different signal-to-noise \changesforfinalver{ratios} (SNR).}
\end{abstract}


\section{Introduction}

\begin{figure*}[htbp]
\centerline{\includegraphics[width=1\textwidth]{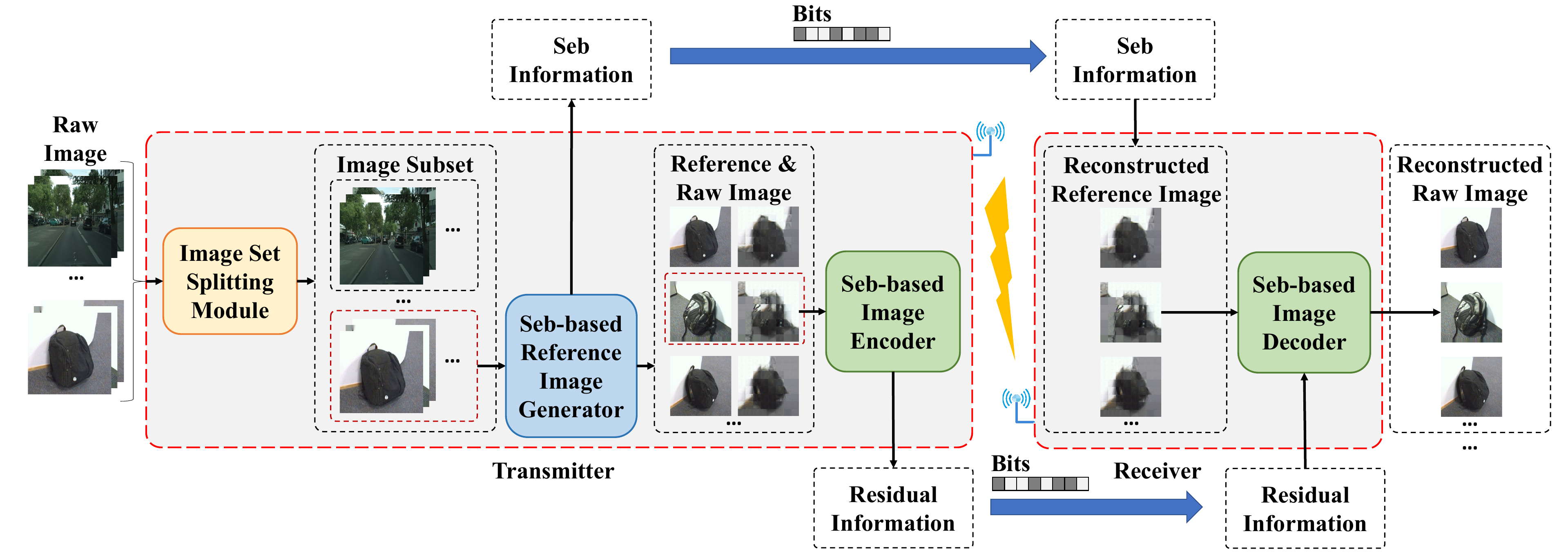}}
\caption{The Seb-based image transmission framework. }
\label{fig:framework}
\end{figure*}


Over the past decades, wireless communications have developed rapidly, where the global mobile traffic is expected to explode up to thousands of exabytes (EB) per month by \Oras{\mbox{2030 \cite{jiang2021}.}}
However, the conventional communication systems based on the classic information theory \rev{focus}
on the bit-level precise transmission while ignoring the meaning of information. With channel capacity and source coding efficiency approaching the Shannon's limit \cite{5, 8}, the conventional communications cannot meet the upsurging communication demands brought by the intelligent services such as Industrial Internet of Things (IIoT), Internet of Vehicles (IoV), and Extended Reality (XR). 

\rev{Recently, benefiting from the great success of deep learning (DL), semantic communications, aiming to successfully convey the meaning of information between transceivers, have experienced vigorous development.}
Most existing works
develop a synchronized knowledge base between the transmitter and the receiver to support the extraction and interpretation of semantics according to the communication intents \cite{fw_w_kb1, fw_w_kb2, fw_w_kb3, fw_w_kb4}, in which knowledge sharing is indispensable. 
However, most of the current works \changesforfinalver{\cite{xu2023, jiang2022, wei2022, hu2023}} take the parameters of the proposed neural network as background knowledge, where the knowledge sharing is implicitly applied in the process of encoding and decoding. Note that the implicit knowledge sharing scheme lacks \Oras{the} awareness of \dlt{the}basic features of semantic representation (e.g. the granularity and efficiency), which can significantly affect the performance of the entire system. To overcome the problem, the concept of semantic base (Seb) is proposed \cite{zhang2022}, \rev{which can be delicately designed with different levels of granularity and similarity to represent semantics w.r.t. intents of communications.} 
The potential of semantic communication\rev{s} is expected to be further \rev{unleashed by properly generating Sebs to form the shared knowledge base between transceivers. However, effective Seb generating schemes have not been developed yet.}
\rev{To} fill the gap, in this paper, a Seb-based semantic communication framework for image transmission is proposed. Taking the complex correlated image sets as input, the proposed framework first splits the image sets according to the correlation level, constructs subsets containing more stably related images for further processing, and generates Sebs to represent the shared semantics as synchronized knowledge. During the transmission, each image is represented by the generated Sebs, and the residual information is further encoded for precise recovery of the image. 
The contributions of this paper are summarized as follows.
\begin{itemize}
    \item \dlt{We propose} A \rev{novel} semantic communication framework with explicit Sebs for image transmission \rev{is proposed, where explicit procedures of Seb generation, transmission, and image semantics representation based on Sebs are included.}
    To the best of our knowledge, this is the first work that utilizes \dlt{the }explicit Sebs \dlt{, including generation, share, and representation,}for image transmission. 
    \item For Seb generation and Seb-based image representation, the {\bf{Seb-based reference image generator}} is \changesforfinalver{dedicatedly} designed.
    \rev{The {\bf{Seb-based image encoder/decoder}} is further proposed to encode/decode the residual information for precise image reconstruction.}
    \rev{A specialized loss function with a gradient approximation mechanism is introduced to enable the E2E training for the entire network that contains a non-derivable clustering algorithm in the {\bf{Seb-based reference image generator}}}.
    \item Extensive experiments \dlt{have been conducted to }validate the effectiveness and the generality of the proposed framework, \rev{which outperforms traditional methods and state-of-the-art DL-based methods by more than \rev{$0.5 - 1.5$} dB w.r.t. peak signal-to-noise ratio (PSNR) under the same signal-to-noise ratio (SNR) and channel bandwidth ratio (CBR). } 
\end{itemize}

\section{Seb-based Image Transmission Framework}

\rev{The Seb-based semantic communication framework is shown in Fig.~\ref{fig:framework}.}
The transmitter includes an {\bf{image set splitting module}}, a {\bf{Seb-Based reference image generator}}, and a {\bf{Seb-based image encoder}}.
\rev{The corresponding Sebs and residual of raw images are extracted and then encoded by a channel encoder for transmission.
At the receiver side, the {\bf{Seb-based image decoder}} reconstructs raw images according to Sebs and their corresponding residual information.}
The detailed \dlt{deployment}\rev{design} of each module will be introduced in the following.


\subsection{Image Set Splitting Module}
\dlt{Considering the changes in potential application scenarios, the correlation levels of the raw images could vary greatly.}
The image set splitting module is employed to split the original image set into several subsets, in which the images are expected to possess higher and more stable correlations that can be better captured by subsequent modules. Denoting $\mathcal{I}\!=\!\{I_1,I_2,...,I_n\},I\!\in\!\mathbb{R}^{C\times H\times W}$ as the original image set, which is composed of $n$ images $I_1,I_2,...,I_n$\rev{, where channels, height\Oras{,} and width of each image are denoted as $C$, $H$\Oras{,} and $W$\Oras{,} respectively.}
We follow the architecture \rev{of contrastive learning \cite{chen}}
to build the image set splitting module, \rev{where}
\begin{equation}
\{\mathcal{I}_1,\mathcal{I}_2,...,\mathcal{I}_J\}  = {f_{\mathrm{cluster}}}({f_{\alpha}}({\mathcal{I},\alpha}),J).
\end{equation}
\dlt{where}$f_{\alpha}(\cdot\Ora{,\alpha})$ denotes the \Ora{projector with its parameters as $\alpha$} that projects each image into latent space, \rev{where similar images are projected into the same cluster.}
The projector is achieved by a Resnet-50 backbone and a projection multilayer perceptron (MLP) \dlt{head }in this paper. $f_{\mathrm{cluster}}(\cdot,J)$ denotes the standard k-means clustering algorithm with $J$ as the number of clusters, which is determined by the clustering inertia. The module outputs $J$ image subsets $\mathcal{I}_j\!=\!\{I_{j(1)},...,I_{j(n_j)}\},j=1,...,J$ with each subset $\mathcal{I}_j$ consisting of $n_j$ images individually. The projector is initialized with the weights in \cite{chen}, and is frozen as an invariant image classifier during training.

\subsection{Seb-based Reference Image Generator}

\begin{figure*}[htbp]
\centerline{\includegraphics[width=1\textwidth]{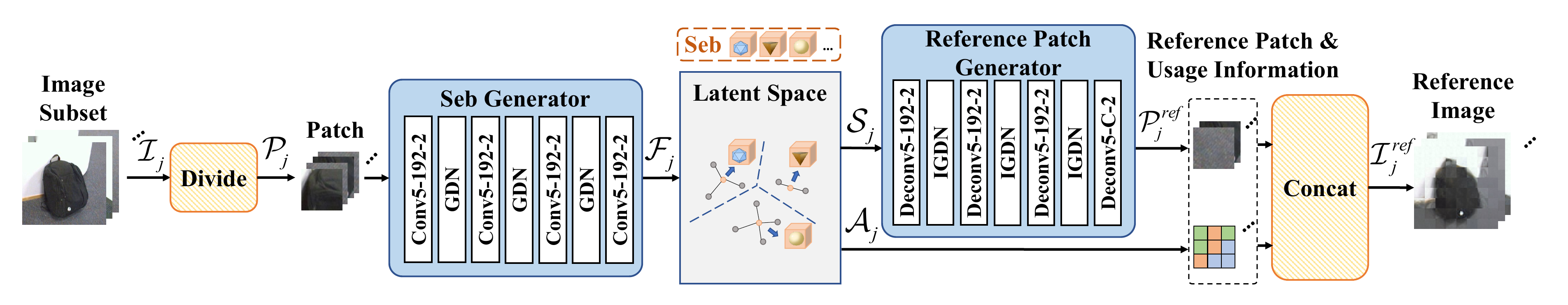}}
\caption{The Seb-based reference image generator. Conv/Deconv5-192-2 represents the convolution/deconvolution operation with $5\times5$ kernel size, 192 output channels, and a stride of 2. GDN/IGDN \cite{balle2016} denotes the nonlinear transform function. The number of Seb $\mathcal{S}_j$ is much smaller than the number of the latent feature of image patches $\mathcal{F}_j$ ($K\ll{n_j}\ast{n_p}$), which supports efficient semantic representation. The whole structure is deployed at the transmitter. }
\label{fig:reference image generator}
\end{figure*}

To achieve \dlt{the procedures of }Seb generation and representation of image\rev{s}\dlt{ semantics based on Sebs}, the Seb-based reference image generator is employed mainly including a Seb generator and a reference patch generator, as shown in Fig.~\ref{fig:reference image generator}. 

\changesforfinalver{In specific, the image subset $\mathcal{I}_j$ is first divided into the patch set $\mathcal{P}_j\!=\!\{P_{j(1)(1)},...,P_{j(1)(n_p)},...,P_{j(n_j)(n_p)}\}$,  where $P\!\in\!\mathbb{R}^{C\times h\times w}$. Each image is divided into $n_p$ patches (e.g., $I_{j(1)}$ corresponds to $\{P_{j(1)(1)},...,P_{j(1)(n_p)}\}$). Parameters $h$ and $w$ denote the patch height and patch width, respectively. }\rev{The operation brings about the reduction of size, and as a result,}
the patches will be mapped into a latent space with lower dimensions so that the semantics will be more precisely represented by Sebs. The granularity of Seb is thus controlled by the patch number and the patch size.

The Seb generator and the reference patch generator are designed based on the autoencoder structure, \changesforfinalver{which} are composed of convolutions and generalized divisive normalization (GDN)/inverse generalized divisive normalization (IGDN) \cite{balle2016} activation functions. The Seb generator is first used to extract semantic features $\mathcal{F}_j=\{F_{j(1)(1)},...,F_{j(1)(n_p)},...,F_{j(n_j)(n_p)}\}$ from the patches $\mathcal{P}_j$, where $F\in\mathbb{R}^{c'\times h'\times w'}$, \rev{with $c',h'\!\!=\!\!\frac{h}{16},$ and $w'\!\!=\!\!\frac{w}{16}$ denoting the dimensions of latent.}

Next, the set of Sebs $\mathcal{S}_j=\{S_{j(1)},...,S_{j(K)}\}$, where $S\in\mathbb{R}^{c'\times h'\times w'}$, and the corresponding usage information $\mathcal{A}_j=\{A_{j(1)(1)},...,A_{j(1)(n_p)},...,A_{j(n_j)(n_p)}\}$, where $A\in\{1,...,K\}$ are generated by a standard clustering algorithm (e.g. K-means) on the semantic features $\mathcal{F}_j$. \Ora{Each Seb $S \in \mathcal{S}_j$ corresponds to the center of each cluster, \rev{where the usage information $A \in \mathcal{A}_j$ records the index of the cluster to which the corresponding semantic feature $F \in \mathcal{F}_j$ belongs.}}
Parameter $K\in\mathbb{Z}^+$ denotes the number of clustering centers, which controls the representation efficiency of Seb\rev{s} \rev{with} $K\ll{n_j}\ast{n_p}$\dlt{ to support efficient semantic representation}. \dlt{Therefore, the semantic features of images are clustered according to their location in latent space, and represented by their clustering centers. }

\rev{To represent images with Sebs,}
the set of reference patches $\mathcal{P}^{ref}_j=\{P^{ref}_{j(1)},...,P^{ref}_{j(K)}\}$, where $P^{ref}\in\mathbb{R}^{C\times h\times w}$ are generated based on Sebs $\mathcal{S}_j$ by mapping the latent into the original space through the reference image generator. The reference image $\mathcal{I}^{ref}_j\!\!\!=\!\!\!\{I^{ref}_{j(1)},...,I^{ref}_{j(n_j)}\},I^{ref}\in\mathbb{R}^{C\times H\times W}$ is \dlt{finally }generated through a concatenating operation under the guidance of Seb usage information $\mathcal{A}_j$ \rev{and the corresponding patches $\mathcal{P}^{ref}_j$}. The mechanism of the {\bf{Seb-based reference image generator}} is described as follows:
\begin{equation}
\begin{aligned}
    &\mathcal{P}_j=f_{\mathrm{divide}}(\mathcal{I}_j),\\
    &(\mathcal{S}_j,\mathcal{A}_j)=f_{\mathrm{cluster}}(f_{\phi}(\mathcal{P}_j;\phi),K),\\
    &\mathcal{I}^{ref}_j=f_{\mathrm{concat}}(f_{\theta}(\mathcal{S}_j;\theta),\mathcal{A}_j),
\end{aligned}
\end{equation}
where $f_{\mathrm{divide}}(\cdot)$, $f_{\mathrm{cluster}}(\cdot)$, and $f_{\mathrm{concat}}(\cdot)$ denote the fore-mentioned dividing, clustering\Oras{,} and concatenating operations\Oras{,} respectively,  $f_{\phi}(\cdot;\phi)$ and  $f_{\theta}(\cdot,\theta)$ denote the corresponding Seb generator and reference patch generator with $\phi$ and $\theta$ stand\Oras{ing} for their trainable parameters.

\Ora{
\rev{Note that Sebs $\mathcal{S}_j$ and the usage information $\mathcal{A}_j$}
need to be synchronized between the transmitter and the receiver. \rev{The wireless channel is modeled as}
\begin{equation}\label{channel}
\begin{aligned}
    &y = hx + z,\\
    &\hat{x} = h^{-1}y = x + \hat{z},
\end{aligned}
\end{equation}
where $x \in \mathbb{C}$ and $y \in \mathbb{C}$ denote the complex symbols of channel input and output, $z \sim \mathcal{CN}(0,\sigma^2)$ denotes the additive white Gaussian noise (AWGN) \rev{with} $\sigma^2$ \rev{as} the average noise power, and $h \in \mathbb{C}$ denotes the channel gain. At the receiver, by estimating the channel state information (CSI), the recovery of the channel input $\hat{x} \in \mathbb{C}$ can be obtained, where $\hat{z} = h^{-1}z$. 

$\mathcal{S}_j$ and $\mathcal{A}_j$ are transmitted after mapping into complex symbols \rev{with} channel coding and modulation, and \rev{are} recovered as $\hat{\mathcal{S}}_j$ and $\hat{\mathcal{A}}_j$ at the receiver.
\rev{The concatenation operation is carried out at the receiver as well to obtain $\hat{\mathcal{I}}^{ref}_j$, where}
\begin{equation}
    \hat{\mathcal{I}}^{ref}_j=f_{\mathrm{concat}}(f_{\theta}(\hat{\mathcal{S}}_j;\theta),\hat{\mathcal{A}}_j).
\end{equation}}

\subsection{Seb-based Image Encoder/Decoder}

\begin{figure}[htbp]
\centerline{\includegraphics[width=0.49\textwidth]{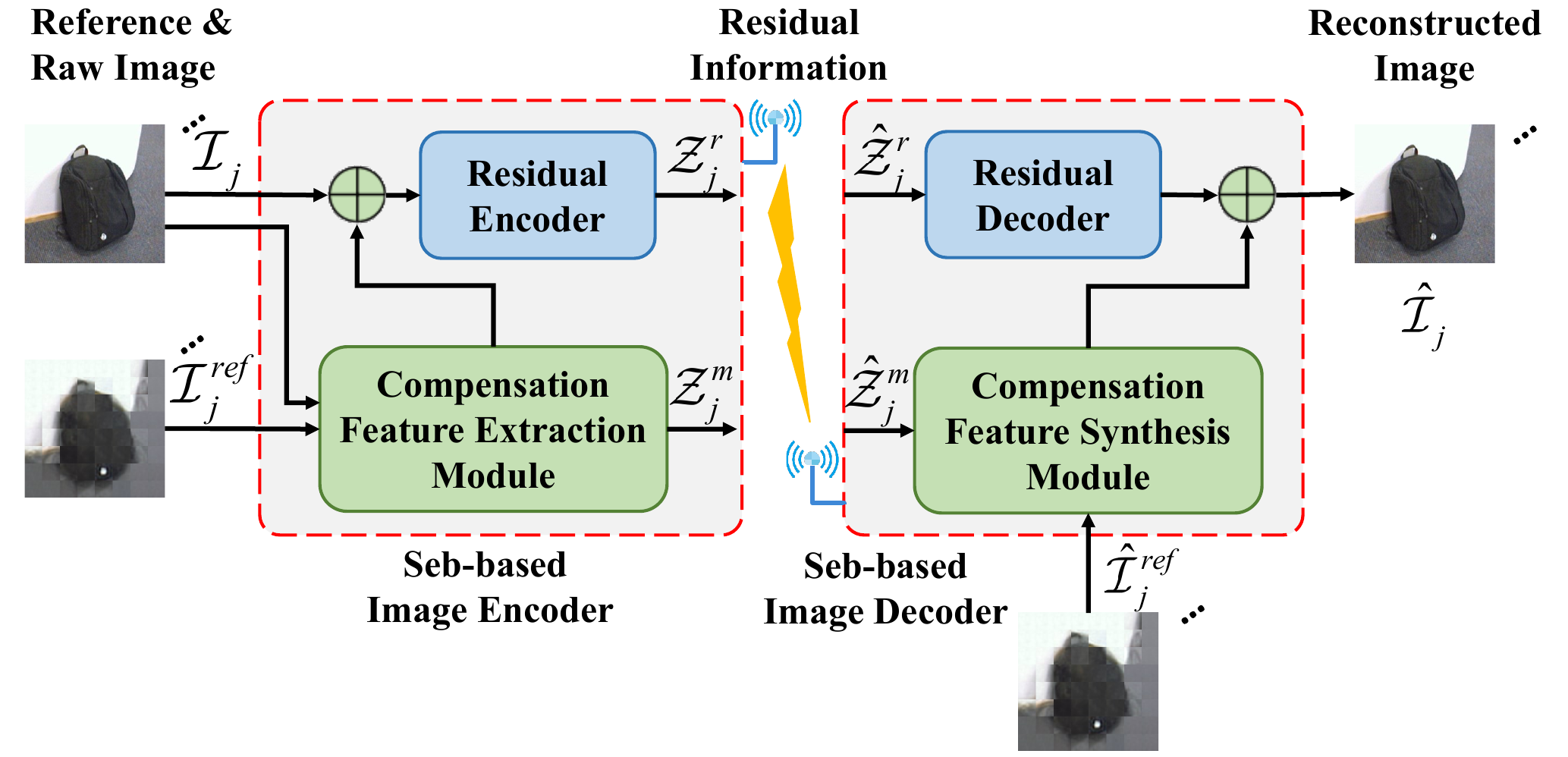}}
\caption{The Seb-based image encoder/decoder.}
\label{fig:image endecoder}
\end{figure}

\Ora{To \rev{achieve precise}
image reconstruction, the {\bf{Seb-based image encoder/decoder}} is proposed to extract residual information based on the reference images $\mathcal{I}^{ref}_j$. It includes a pair of compensation feature extraction/synthesis module \dlt{as well as }\rev{and} a pair of residual encoder/decoder, as shown in Fig.~\ref{fig:image endecoder}. 
\footnote{
We follow the \rev{structure of} E2E deep video compression scheme \cite{lu2019} to build the proposed Seb-based image encoder/decoder in this paper.
\rev{The structure of the Seb-based image encoder/decoder will be further optimized in the future journal manuscript.}
}}

In the encoding process, the compensation feature extraction module takes the raw images $\mathcal{I}_j$ and the generated reference images $\mathcal{I}^{ref}_j$ as inputs, and uses a CNN-based optical flow net \cite{ranjan} and a corresponding encoder to extract the compensation features $\mathcal{Z}^m_j$\dlt{ to generate the predictions toward the raw images} \rev{that corresponds to the predictions to the raw images}. The residual encoder further compresses the residual parts between the raw images and their predictions into \Ora{the residual features} $\mathcal{Z}^r_j$. 
\Ora{The features $\mathcal{Z}^m_j$ and $\mathcal{Z}^r_j$ need to be transmitted \dlt{similar to $\mathcal{S}_j$ and $\mathcal{A}_j$}\rev{through the wireless channel as depicted in (\ref{channel})}, and be recovered as $\hat{\mathcal{Z}}^m_j$ and $\hat{\mathcal{Z}}^r_j$ at the receiver. }
In the decoding process, the residual parts and the predictions of the images are obtained from the recovered $\mathcal{\hat{Z}}^m$ and $\mathcal{\hat{Z}}^r$ by the residual decoder and the compensation feature synthesis module\Oras{,} respectively. Then the reconstructed image set $\hat{\mathcal{I}}_j$ is obtained through a summation operation.


\section{Problem Formulation and Training Strategy}
The key components of the proposed framework are trained in an E2E manner, and are jointly optimized based on rate-distortion trade-off through a specialized loss function. 

\subsection{Rate-Distortion Optimization}

The goal of \dlt{our }\rev{the proposed} Seb-based image transmission framework is to minimize the number of \rev{transmitted} bits\dlt{used to transmit the image set}, while making the distortion between the raw image set $\mathcal{I}$ and the reconstructed image set $\hat{\mathcal{I}}$ as small as possible. This is a typical rate-distortion optimization problem \rev{that is modeled as}
 \begin{equation}
L_{RD}  = \lambda D(\mathcal{I},\hat{\mathcal{I}})+(R(\mathcal{S})+R(\mathcal{A})+R(\mathcal{Z}^m)+R(\mathcal{Z}^r)).
\end{equation}
\dlt{where }
$D(\mathcal{I},\hat{\mathcal{I}})$ denotes the distortion between the raw and reconstructed image set, which is represented by the mean square error (MSE) in training. $R(\mathcal{S}), R(\mathcal{A}), R(\mathcal{Z}^m), R(\mathcal{Z}^r)$ denote the bitrate corresponding to \Ora{the Seb, the Seb usage information, as well as the compensation and the residual features\Oras{,} respectively}. \rev{$\lambda$ is a hyper-parameter that controls the trade-off between rate and distortion. \Oras{Higher $\lambda$ frameworks tend to consume more resources for less distortion.}}

In specific, the bitrate of Seb for the image set $\mathcal{I}$ is approximated as 
\begin{equation}
R(\mathcal{S})=\sum\limits_{\mathcal{S}_j \in \mathcal{S}}R(\mathcal{S}_j) = \sum\limits_{\mathcal{S}_j \in \mathcal{S}} \sum\limits_{P^{ref}\in\mathcal{P}^{ref}_j}R(P^{ref}),
\end{equation}
where $R(P^{ref})$ denotes the transmitting cost of \changesforfinalver{$P^{ref}$}.
The bitrate of Seb usage information $\mathcal{A}$ is calculated as 
\begin{equation}
R(\mathcal{A})\!\!=\!\!\sum\limits_{\mathcal{A}_j \in \mathcal{A}}\!\!R(\mathcal{A}_j)\!\!=\!\!\sum\limits_{\mathcal{A}_j \in \mathcal{A}}\!\!\sum\limits_{A \in \mathcal{A}_j}\!\!R(A)\!=\!n\times\!n_p\!\times\!\log_2K,
\label{eq:usage info}
\end{equation}
where $n_p \ast \log_2K$ denotes the constant cost of each image $I\in\mathcal{I}$. The bitrate $R(\mathcal{Z}^m)$ and $R(\mathcal{Z}^r)$ should be measured as the entropy of the corresponding latent representation symbols. In this paper, we use the entropy model in \cite{balle2018} to estimate $R(\mathcal{Z}^m)$ and $R(\mathcal{Z}^r)$\Oras{,}\Ora{ respectively,
which can be expressed as follows, 
\begin{equation}
\begin{aligned}
    &R(\mathcal{Z}^m)=-\log_2p(\mathcal{Z}^m|\mathcal{I},\delta_m),\\
    &R(\mathcal{Z}^r)=-\log_2p(\mathcal{Z}^r|\delta_r),
\end{aligned}
\end{equation} 
where $\delta_m$ and $\delta_r$ denote the parameters of the parametric and non-parametric entropy model\Oras{,} respectively, discriminated by whether they directly depend on the input images. $p(\mathcal{Z}^m|\mathcal{I},\delta_m)$ and $p(\mathcal{Z}^r|\delta_r)$ denote the corresponding estimated probability of $\mathcal{Z}^m$ and $\mathcal{Z}^r$}. 

\subsection{Gradient Approximation for Clustering Operation} 

Note that \changesforfinalver{the Seb-based reference image generator employs a clustering algorithm,} the module before the clustering algorithm (i.e., the Seb generator) cannot be updated in the back-propagation. Inspired by \cite{aaron2018}, in this paper, we directly copy the gradient of each of the reference patch generator's input $S_{i}\in\mathcal{S}$ back to the corresponding outputs of the Seb generator $\mathcal{F}_{i}=\{F_{i}|f_{\mathrm{cluster}}(F_{i})=S_{i},F_{i}\in\mathcal{F}\}$.
\rev{However,}
\rev{t}he direct copy operation \dlt{will }make\rev{s} the clustering operation not constrained by E2E loss, resulting in an arbitrarily grown latent space. To tackle the problem, the \rev{$L_2$ regulation is utilized}
to move each $F_{i}$ towards $S_{i}$, 
\begin{equation}
\begin{aligned}
    L_{Reg}=Reg(\mathcal{F},\mathrm{sg}(\mathcal{S}))=\sum\limits_{F_i \in \mathcal{F}}\sum\limits_{S_i \in \mathcal{S}}||F_i-\mathrm{sg}(S_i)||^2_2,\\
\end{aligned}
\label{eq:regloss}
\end{equation}
where $\mathrm{sg}(\cdot)$ represents the stop gradient operator that constraining $\mathcal{S}$ to not be directly moved.

As a result, the total loss function can be expressed \changesforfinalver{as}
\begin{equation}
\begin{aligned}
    L&=L_{RD}+\beta L_{Reg},\\
\end{aligned}
\end{equation}
where $\beta$ controls the weight of the \Ora{regulation} loss. We use $\beta=1$ in our experiments.

\addtolength{\topmargin}{0.02in}
\section{Experimental Results}
\Ora{In this section, we conduct extensive experiments to validate the effectiveness and generality of \rev{the} proposed framework. The validation methodology is first provided, including the construction of training and testing datasets, the \rev{parameters setting},
the choice of baselines, \rev{and} the performance metrics for comparison. \rev{Then, the experimental results are further discussed to illustrate the performance of the proposed method.}
}

\subsection{Methodology}
\rev{As for training}, a training set consisting of 50,000 images is sampled from the ImageNet training dataset \cite{imagenet}, with each image being enhanced (randomly resized and cropped) into 16 augmentations with 256 $\times$ 256 resolution as image subsets with a certain level of correlation during training. 
The Adam optimizer \cite{adam} is used by setting $\beta_1$ \rev{and $\beta_2$ as $0.9$ and $0.999$\Oras{,} respectively}.
The framework is trained at the learning rate of $10^{-4}$ and $10^{-5}$ each for one epoch, and continued at the learning rate of $10^{-6}$ for three epochs until convergence.

\Oras{To illustrate the advantages of the proposed framework, during validation,} 
we construct a Mixed dataset by mixing samples from the Cityscapes' test set \cite{city} and the UVG \mbox{dataset \cite{uvg}}. \changesforfinalver{These two} datasets are composed of photos of urban street scenes and frames of video sequences, respectively, containing samples with distinct distributions. Specifically, the Mixed dataset consists of 500 images with 100 sampled from an UVG sequence, others sampled from Cityscapes' test set, and all images are cropped into 1024 $\times$ 1920 resolution.
\Oras{The Mixed dataset denotes a representative use case of the proposed framework, where images with different levels of correlations and characteristics need to be transmitted.}

We specify the Seb representation efficiency parameter $K=\lfloor\frac{n_j\ast{n_p}}{25}\rfloor$ \rev{to make}
the number of Sebs \rev{with} fixed proportion to the number of images in each subset, where $\lfloor\cdot\rfloor$ denotes the floor operation. The patch height and \changesforfinalver{width} are set as $h=w=32$. 
We compare the proposed framework with the following baselines, 
\begin{itemize}
    \item The wide-used engineered image compression codec JPEG2000 \cite{jp2} \dlt{(marked as ``JPEG2000'')}\rev{as the representative baseline for traditional methods}.
    \item The DL-based compression scheme \cite{balle2018} (\rev{denoted} as ``DL w/o cor\rev{r}.''). The scheme utilizes an autoencoder structure for the compression and reconstruction of images, which each image be processed independently. The correlations among \dlt{the }images are ignored.
    \item The DL-based compression scheme \cite{lu2019} (\rev{denoted} as ``DL w/ cor\rev{r}.''). The scheme achieves compression based on image pairs. A prediction of the target image is made based on the reference image, and then the autoencoders are used to compress the prediction and the residual parts\Oras{,} respectively.\dlt{The reference image is typically a reconstruction of the other image in the dataset.} The \rpl{corerelations}{correlations} among \dlt{the }images are \dlt{relatively }strict.
\end{itemize}

\rev{Note} that the same autoencoder structure is used in \cite{balle2018, lu2019} and the proposed framework. Therefore, the comparison with these two \rpl{schemes}{baselines} avoids the impact caused by the difference in basic network structure to a certain extent, which can reflect the effectiveness of the \rev{proposed} Seb-based method \rev{more accurately}. 

Without loss of generality, the experiments are taken over the AWGN channel, with each scheme combined with an ideal capacity-achieving channel code for transmission. We test the performance of each scheme under different signal-to-noise ratio (SNR) and channel bandwidth ratio (CBR) \cite{bourt, wang2022} conditions, \changesforfinalver{which reflect} the channel condition and the overall coding rate\Oras{,} respectively. In specific, CBR is defined as the ratio between the number of channel input symbols (e.g. the number of \Ora{transmitted} symbols of $\mathcal{S},\mathcal{A},\mathcal{Z}^m$, and $\mathcal{Z}^r$ for the proposed framework) and the number of source image symbols (e.g. $3 \times 1024 \times 1920$ for images in $\mathbb{R}^{3\times 1024\times 1920}$). In this case, we can derive the function of SNR and bit-per-pixel (BPP) of the above schemes under a given CBR, which by first the channel capacity $C$ is obtained as
\begin{equation}
    C = \frac{\mathrm{BPP}}{3 \times \mathrm{CBR}},
\end{equation}
and then SNR is obtained according to the Shannon channel capacity formula,
\begin{equation}
    C=\log_2(1+\mathrm{SNR}).
\end{equation}
Peak signal-to-noise ratio (PSNR) and multi-scale structural similarity (MS-SSIM) \cite{wang2003} are used to measure the quality of image reconstruction.

\subsection{Results and Analysis}

\begin{figure}[htbp]
\centering
\subfigure[]{\includegraphics[width=0.3\textwidth]{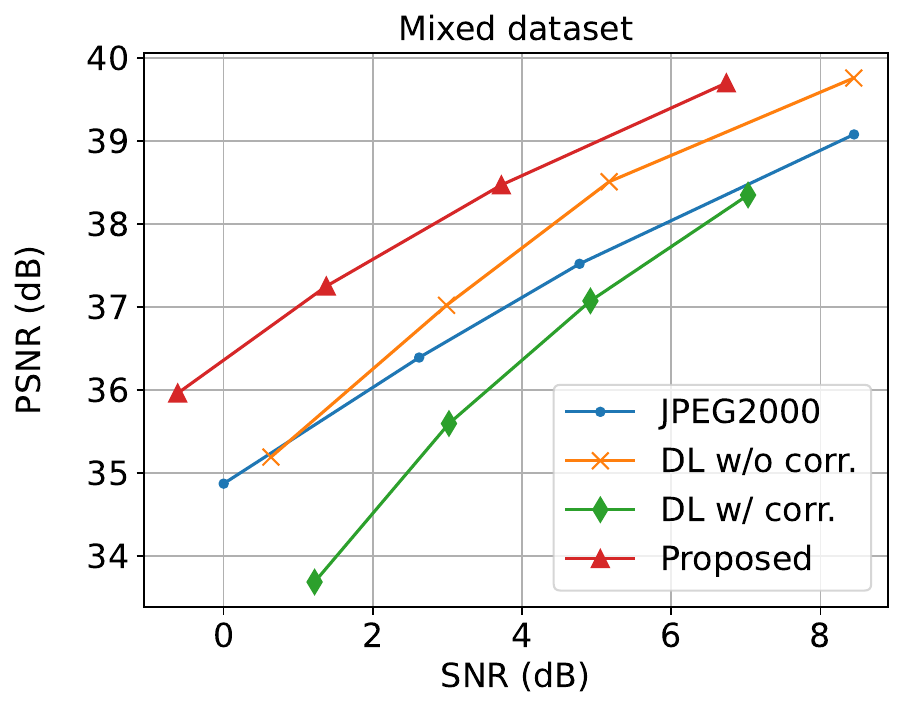}\label{fig:snr_psnr_mix}}\vspace{-0mm}
\subfigure[]{\includegraphics[width=0.3\textwidth]{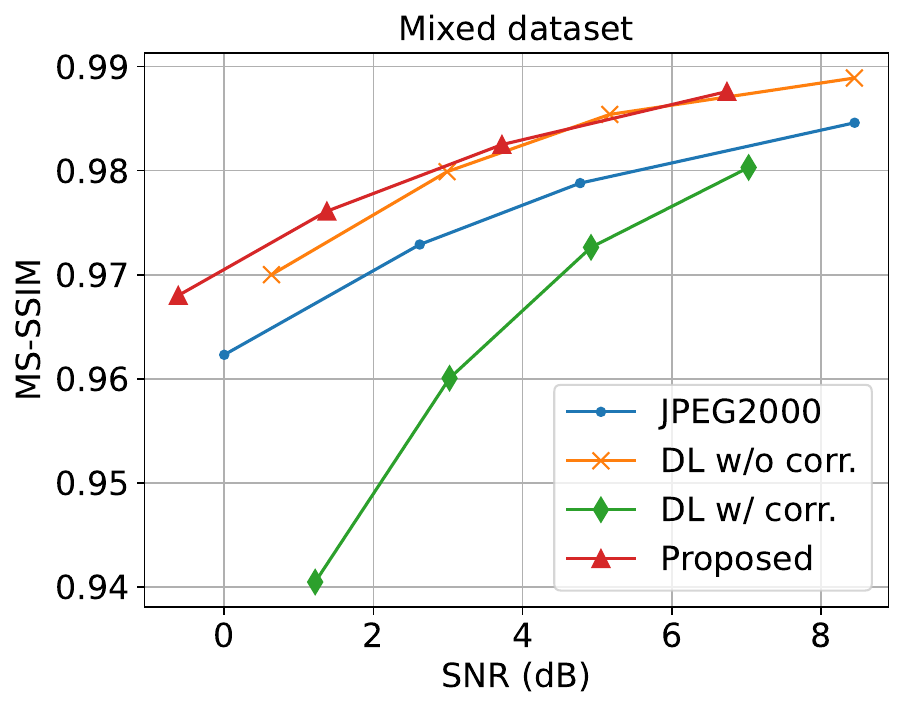}\label{fig:snr_msssim_mix}}\vspace{-0mm}
\caption{SNR-distortion curves on the Mixed dataset with CBR = 1/30. \com{(The other two points, and performance with other lambda of the updated version is still on running.)}}
\label{fig:performance on mixed dataset}
\end{figure}

Fig.~\ref{fig:performance on mixed dataset} shows the PSNR and MS-SSIM results under different SNR conditions on the mixed dataset with CBR = 1/30. The proposed framework outperforms the baselines in general, achieving about $0.5 - 1.5$ dB gain when measured in PSNR.
\changesforfinalver{Notably, the scheme \cite{lu2019} (DL w/ corr.) shows the worst performance due to the statistical difference between the training and testing data, which makes it unable to perform effective correlation information extraction.}
For the proposed framework, the performance gain is more significant under low SNR conditions, whereas the other DL-based schemes show more severe performance deterioration. 
\Ora{This is because of the introduction of the Seb-based image representation mechanism, which supports an efficient recovery of the images' semantics while using limited communication resources. 
\Oras{Besides, the framework also achieves similar or better performance w.r.t. MS-SSIM, as shown in Fig.~\ref{fig:snr_msssim_mix}.}}

\begin{figure}[htbp]
\centerline{\includegraphics[width=0.3\textwidth]{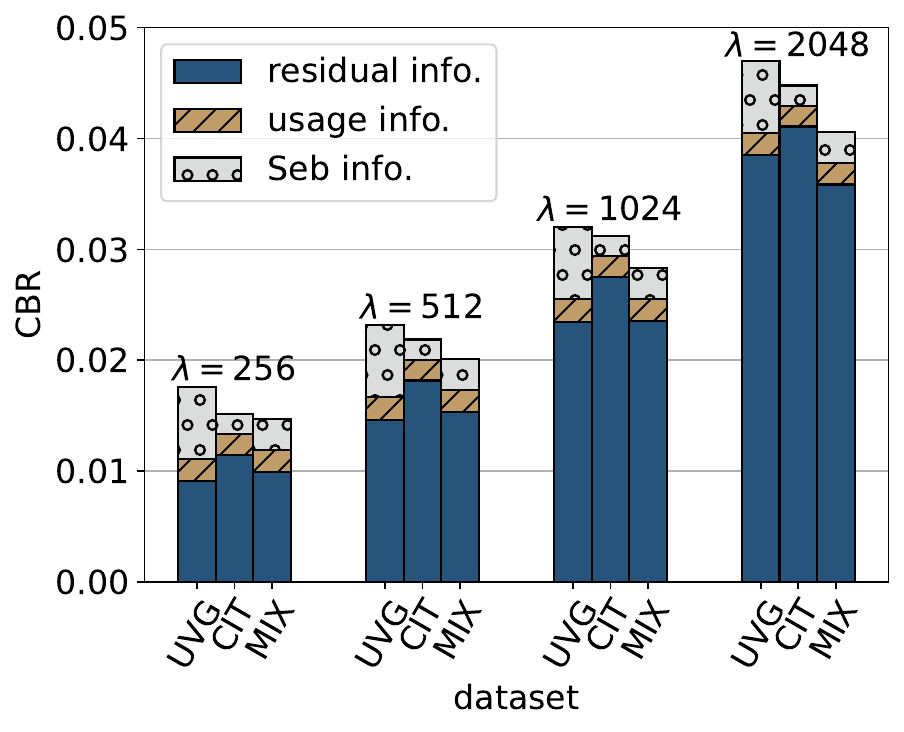}}\vspace{-2mm}
\caption{The average CBR consumption of Sebs, usage information, and residual information for one image under different $\lambda$ on the UVG, Cityscapes, and the Mixed datasets with SNR = 5.
}
\label{fig:lambda_cbr}
\end{figure}

Fig.~\ref{fig:lambda_cbr} shows the average CBR consumption of the proposed framework corresponding to Sebs, usage information, and residual information under different $\lambda$ \rev{($\lambda=\{256,512,1024,2048$\})} and datasets with SNR = 5. It can be observed that the proportion of Seb information varies significantly under different settings. The UVG dataset consumes the highest proportion of Seb information, followed by the Mixed and the Cityscapes datasets\Oras{,} respectively\Oras{,} under the same $\lambda$ conditions. The result is in accordance with the correlation level of datasets that Seb carries most of the information for a strongly correlated dataset, leaving a small amount of residual information. 
For frameworks under different $\lambda$, a relatively fixed amount of Seb information, which depends on the characteristics (e.g. the complexity of image textures) of datasets, is consumed to support the recovery of images' semantics, and more residual information \Oras{is consumed by} frameworks with higher $\lambda$ \Oras{to satisfy the requirement of higher quality of image recovery}. 
In addition, the CBR consumption of usage information is consistent with the result in (\ref{eq:usage info}), which is nearly invariant under specific $n_h,n_w$, and $K$ settings.


\begin{figure}[htbp]
\centering
\subfigure[Original]{\includegraphics[width=0.22\textwidth]{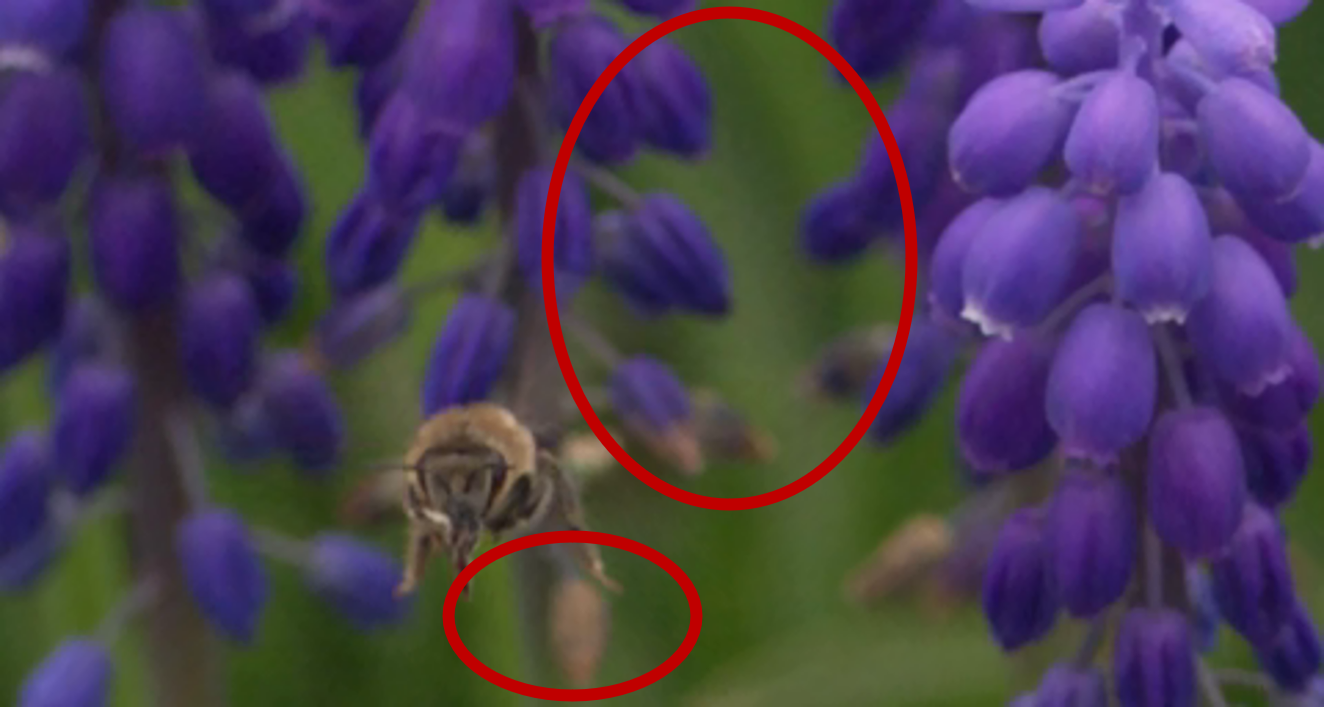}\label{fig:ori}}
\subfigure[JPEG2000 0.0162CBR]{\includegraphics[width=0.22\textwidth]{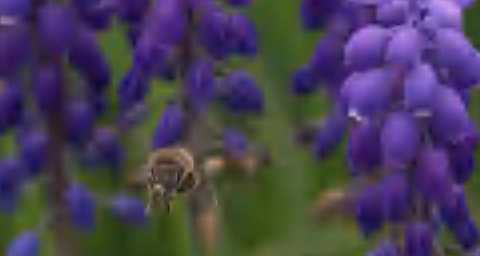}\label{fig:jp2_recon}}
\subfigure[DL w/o corr. 0.0221CBR]{\includegraphics[width=0.22\textwidth]{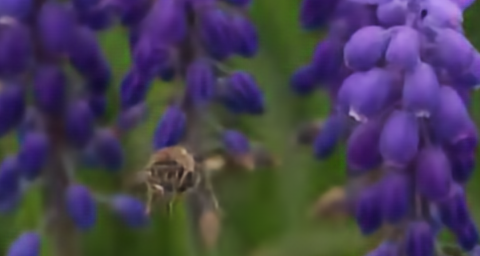}\label{fig:balle_recon}}
\subfigure[DL w/ corr. 0.0113CBR]{\includegraphics[width=0.22\textwidth]{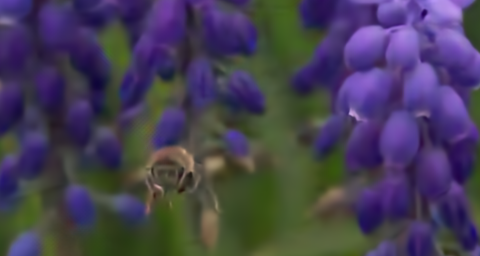}\label{fig:dvc_recon}}
\subfigure[Proposed 0.0075CBR]{\includegraphics[width=0.22\textwidth]{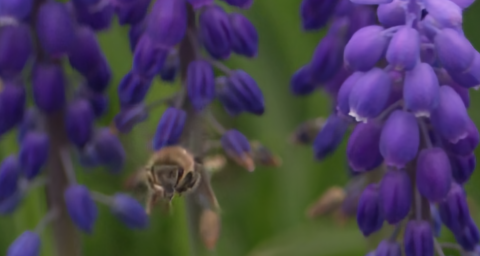}\label{fig:proposed_recon}}
\subfigure[Seb representation]{\includegraphics[width=0.22\textwidth]{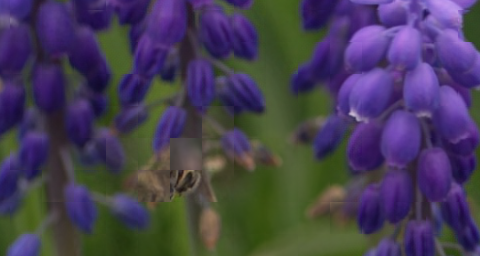}\label{fig:proposed_seb}}
\caption{Examples of reconstructions by different transmission schemes for low CBR values with SNR = 5. }
\label{fig:reconstruction samples}
\end{figure}


Fig.~\ref{fig:reconstruction samples} shows the reconstruction samples of different schemes. Compared to the relatively dynamic region (honeybee), the static region (flower) is more precisely represented by Seb as shown in Fig.~\ref{fig:proposed_seb}. The reason is that the static patterns are more densely distributed when mapping into the latent space, thus allowing for a better clustering result. Moreover, compared to the baselines, the proposed framework presents higher reconstruction quality (more precise color and details as shown in the oval region) with a significantly lower amount of transmitted data. Since the overall region of the image is effectively reconstructed by Sebs, only a small amount of residual is required to restore details, which is a typical use case for the proposed framework. 

\section{Conclusion}
In this paper, we propose \rpl{the}{a} Seb-based image transmission framework, where \rev{common} knowledge \rev{between the transmitter and the receiver} is explicitly formed and shared in the form of Sebs. The framework includes a Seb-based reference image generator for Seb generation and Seb-based image representation, and a Seb-based image encoder/decoder to encode/decode the residual information for precise image reconstruction. A specialized loss function is \dlt{also }introduced to solve the non-derivable problem. Experimental results show that the proposed framework outperforms \changesforfinalver{baselines} \rpl{ in processing complex correlated images}{under all tested channel conditions}. Future work would focus on further refining the generation and representation mechanism of Seb\rev{s}.

\end{document}